\documentclass[nofootinbib,10pt, twocolumn]{revtex4-1}
\usepackage[utf8]{inputenc}
\usepackage[T1]{fontenc}
\usepackage[paper=a4paper,left=25mm,right=25mm,top=25mm,bottom=20mm]{geometry}
\usepackage[english]{babel}
\usepackage{setspace}
\usepackage{physics}
\usepackage{mhchem}
\usepackage{mathtools}
\usepackage{bm}
\usepackage{graphicx}
\usepackage[stable]{footmisc}
\usepackage{xcolor}
\usepackage{soul}
\usepackage{hyperref} 
\hypersetup{colorlinks=false, citebordercolor=green}

\date{\today}

\begin{document}
\title{Continuum limit of bipartite lattices - The SSH model}

\author{Fotios K. Diakonos$^1$}
\email{fdiakono@phys.uoa.gr}
\author{Peter Schmelcher $^{2,3}$}
\email{peter.schmelcher@uni-hamburg.de}
\affiliation{$^1$Department of Physics, National and Kapodistrian University of Athens, GR-15784 Athens Greece}
\affiliation{$^2$Center for Optical Quantum Technologies, University of Hamburg, Luruper Chaussee 149, 22761 Hamburg, Germany}
\affiliation{$^3$The Hamburg Centre for Ultrafast Imaging, University of Hamburg, Luruper Chaussee 149, 22761 Hamburg, Germany}

\begin{abstract}
We present a continuous non-local model that faithfully replicates the rich topological and spectral features of the Su-Schrieffer-Heeger (SSH) model. Remarkably, our model shares the SSH models' bulk energy spectrum, eigenstates, and Zak phase — hallmarks of its topological character — while introducing a tunable length-scale $a$ quantifying non-locality. This parameter allows for a controlled interpolation between non-local and local regimes. Furthermore, for a specific value of $a$ the exact spectral equivalence to the discrete SSH model is established. Distinct from previous continuous analogues based on Schr\"{o}dinger or Dirac-type Hamiltonians, our approach maintains chiral symmetry, does not require an external potential and features periodic energy bands. On finite domains, the model supports a flat band with zero energy formed by a countable infinite set of exponentially localized zero-energy edge states of topological origin.  Beyond SSH, our method lays the foundation for constructing non-local, continuous analogues of a wide class of bipartite and multipartite lattices, opening new paths for theoretical exploration and new challenges for experimental realization in topological quantum matter.

\end{abstract}

\maketitle

\section{Introduction}

In recent decades, the Su-Schrieffer-Heeger (SSH) model—originally introduced to describe polyacetylene's electronic structure \cite{Su1979}—has become a paradigm for understanding topological properties in connection to periodic solid state systems. Despite its simplicity, the model elegantly captures the interplay between symmetry, topology, and the occurrence of edge states, constituting a typical example for the accomplishment of the \textit{bulk-boundary correspondence} principle \cite{Asboth2016,Hasan2010,Cooper2019}. The discrete one-dimensional SSH chain consists of unit cells with two sublattice sites (A and B), featuring an intra-cell hopping amplitude $v$ and an inter-cell amplitude $w$. An infinite chain exhibits a topologically nontrivial phase when $\vert w \vert > \vert v \vert$ and a trivial phase when $\vert v \vert > \vert w \vert$. This distinction is characterized by the Zak phase \cite{Zak1989}, a Berry phase integral across the Brillouin zone that takes values of $1$ (topological phase) or $0$ (normal phase). Under chiral symmetry, the Zak phase is closely related to a winding number in the Hamiltonians' parameter space, establishing a winding number classification for 1D chirally symmetric systems \cite{Asboth2016,Chiu2016}.

Recent advances have spurred interest in extending discrete models like SSH to continuous frameworks, facilitating exploration of topological phenomena in e.g. photonic crystals \cite{Naz2018,Wang2022,Tang2022,Gupta2022}, acoustic metamaterials \cite{Li2018,Zheng2019,Coutant2022,Liu2023}, and quantum field theories \cite{Dong2019,Salvo2024}. 
Among the same lines, several works have explored 1D continuous SSH analogues by modifying geometric or dynamical properties, focusing on phenomenological similarities or/and discrepancies 
\cite{Coutant2021,Lustig2017,Quemerais1993,Zoli04,Goren2018}.
Experimental platforms such as optical superlattices in cold-atom physics enable studies of topological properties in continuous systems \cite{Dalibard2018,Zhang2018,Cooper2019,Katsaris2024}.

In general, the spatially continuous counterparts of the SSH model can be divided into two categories: (i) models based on Schr\"{o}dinger-type Hamiltonians and (ii) models based on Dirac-type Hamiltonians. In the former case, the usual continuous SSH counterpart contains an external potential term which forms a double-well potential landscape within each cell to generate the different intra- and inter-cell hoppings. This potential usually modifies the energy spectrum, allowing for tunneling processes which correspond to next-to-nearest-neighbor hopping. Consequently, this leads to the breakup of chiral symmetry. Then, a fine adjustment in the potential parameters is needed to approximately restore this symmetry and generate the topological properties of the corresponding SSH model \cite{Dalibard2018,Katsaris2024}. 


Furthermore, the bulk spectrum is different from that of the SSH model in these Schr\"{o}dinger-type continuous models since they provide infinitely many energy bands missing in the discrete counterpart. Thus, no clear one-to-one relationship, both in pure spectral and topological properties, between the discrete SSH model and its continuous Schr\"{o}dinger-type counterparts can be established. In addition, a recent demonstration of the nontrivial relationship between bipartite lattices and their continuous counterparts \cite{Shapiro2022a} shows that two homotopically equivalent 1D continuous Schr\"{o}dinger-type Hamiltonians lead to topologically distinct SSH chains in the tight-binding limit. This suggests that the SSH model's topology in this case may arise from sublattice symmetry emergent through discretization. Intriguingly, dimensionality plays a crucial role: 2D continuous integer quantum Hall systems of Schr\"{o}dinger type share topological invariants with their discrete counterparts \cite{Shapiro2022b}, in clear deviation from the 1D case.

In contrast to the Schr\"{o}dinger-type models the continuous-space models based on Dirac-type Hamiltonians do not need an external potential to generate spectral and topological characteristics of the discrete SSH model. As a consequence, exact chiral symmetry is present in these models. A representative of this class is the Jackiw-Rebbi model \cite{Jackiw1976}, introduced even before the discrete SSH model. This model serves as a benchmark of a spatially continuous model exhibiting chiral symmetry, symmetric continuous spectrum, and localized zero-energy states, features shared with the discrete SSH model. However, unlike SSH, the bulk spectrum of the Jackiw-Rebbi model does not possess a finite Brillouin zone, a property which is common among continuous-space 1D models emulating SSH phenomenology, based on Dirac-type Hamiltonians.

In the present work, we investigate the possiblity to formulate an one-dimensional model defined in continuous space which combines the existence of finite Brillouin zone (infinite system) with the emergence of localized topologically protected edge modes (finite system). To this end, we introduce a continuous 1D model that reproduces the infinite discrete SSH models' bulk spectrum while supporting chiral zero-energy edge states when restricted to finite space. It is non-local, with a parameter specifying the spatial scale of non-locality in the model. This parameter controls the period of the Brillouin zone as well as the localization length of the emerging edge states. Within this model it is possible to  study the transition from non-local to local description, observing at the same time the associated modifications in the bulk spectrum and the corresponding topological properties. In addition to being non-local, the model also differs from previously proposed Schr\"{o}dinger-type models since it does not contain external potential terms. As a consequence, there are no additional energy bands, and the chiral symmetry is exactly preserved. Furthermore, the model differs from Dirac-type Hamiltonians because it is non-local, which turns out to be crucial for obtaining energy bands in the bulk, which are periodic functions of the wave vector $k$. When formulated on a finite space interval, the associated spectrum contains a flat band at zero energy involving an infinite countable set of exponentially localized edge states. This property suggests a reconsideration of bulk-boundary correspondence in the case of spatially continuous systems with non-trivial topology. 

This work is structured as follows. In Section II, we present the non-local continuous SSH model and analyze its symmetries, spectral properties, and bulk topology. In Section III, we introduce a systematic approach to obtain local approximations of the non-local model. We also describe how the system transitions from non-local to local behavior, focusing on the resulting changes in the bulk spectrum and related topological features. In Section IV, we investigate the finite-size formulation of the introduced non-local model and reveal the emergence of a zero-energy flat mini band consisting of a countable infinite set of localized and topologically protected edge states.  Finally, Section V provides our conclusions and perspectives.

\section{Non-local continuous SSH: spectrum and bulk topology}

\begin{figure}[htp]
\begin{center}
\includegraphics[width=0.5\textwidth]{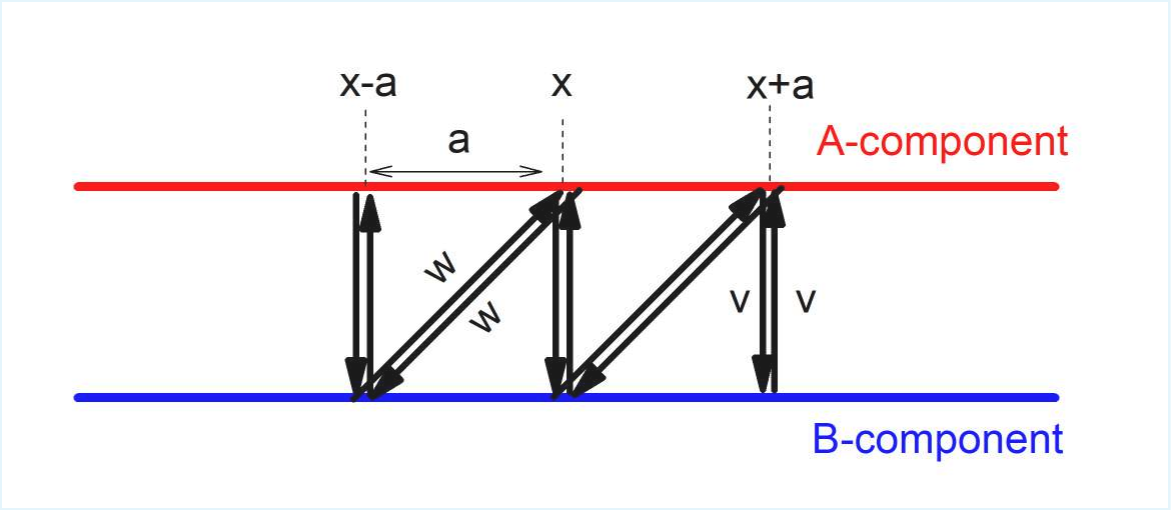}
\caption{Sketch of the continuous one-dimensional SSH model proposed in the current work. The value of the wave field at $x$ is depending on the corresponding values at $x \pm a$ in the specific manner shown here.}
\label{fig:fig1}
\end{center}
\end{figure}

Let us introduce the $2 \times 2$ Dirac-type bulk Hamiltonian $\mathbf{H}$, defined as:
\begin{equation}
\mathbf{H}=\left(w \cos (a \mathbf{p})+v\right) \bm{\sigma}_x + w \sin(a \mathbf{p}) \bm{\sigma}_y
\label{eq:eq1}
\end{equation}
where $\mathbf{p}={\hbar \over i}{d \over dx}$ is the momentum operator in one spatial dimension and $\bm{\sigma}_i$ (with $i=x,y,z$) are the Pauli matrices. Without loss of generality, the parameters $w$, $v$ are assumed to be real while $a$ is real and positive. The Hamiltonian $\mathbf{H}$ fulfills the following relations:
\begin{equation}
\{\mathbf{H},\bm{\sigma}_z\}=0~~~~,~~~~~[\mathbf{H},\bm{\sigma}_x \mathbf{\Pi}(x)]=0
\label{eq:eq2}
\end{equation}
with $\mathbf{\Pi}(x)$ being the $x$-parity operator. The anticommutation relation occurring firstly in Eq.~(\ref{eq:eq2}) expresses the chiral symmetry of $\mathbf{H}$, while the subsequent commutation relation expresses the generalized parity symmetry in this spatially continuous model. Thus, the Hamiltonian $\mathbf{H}$ in Eq.~(\ref{eq:eq1}) possesses the characteristic symmetries of the standard discrete SSH model. One can write $\mathbf{H}$ using the basis of momentum-operator eigenstates $\vert  k \rangle$ satisfying $\mathbf{p} \vert k \rangle = \hbar k \vert k \rangle$ to obtain the form:
\begin{equation}
    \mathbf{H}(k) = \begin{pmatrix}
        0 & w e^{-\mathfrak{i} a k} + v \\
        w e^{\mathfrak{i} a k} + v & 0
    \end{pmatrix}
    \label{eq:eq3}
\end{equation}
which is identical to the discrete SSH bulk Hamiltonian for $a=1$. In fact, the matrix in Eq.~(\ref{eq:eq3}) can be written in terms of the translation operator $\mathbf{T}(a)=\displaystyle{e^{a {d \over dx}}}$ -- with the action $\mathbf{T}(a) f(x)=f(x+a)$ on a function $f(x)$ -- as follows:
\begin{equation}
    \mathbf{H}=\begin{pmatrix}
        0 & w ~\mathbf{T}(-a) + v \\ w~ \mathbf{T}(a) + v & 0   \end{pmatrix}
       \label{eq:eq4} 
\end{equation}
The matrix $\mathbf{H}$, as defined in Eqs.~(\ref{eq:eq1}, \ref{eq:eq4}), acts on spinors of the form:
\begin{equation}
  \mathbf{\Psi}(x)=\begin{pmatrix} \Psi_A (x) \\ \Psi_B(x) \end{pmatrix}
  \label{eq:eq5}
\end{equation}
The spectral properties of $\mathbf{H}$ are determined by the eigensystem:
\begin{equation}
    \mathbf{H} \mathbf{\Psi}(x) = E \mathbf{\Psi}(x)
    \label{eq:eq6}
\end{equation}
where $E$ is an eigenvalue of $\mathbf{H}$ and $\mathbf{\Psi}(x)$ the corresponding eigenvector. Using the representation in Eq.~(\ref{eq:eq4}) for $\mathbf{H}$ the eigensystem in Eq.~(\ref{eq:eq6}) leads to the set of equations:
\begin{eqnarray}
    w \Psi_B(x-a) + v \Psi_B(x) &=& E \Psi_A(x) \nonumber \\
    v \Psi_A(x) + w \Psi_A(x+a) &=& E \Psi_B(x)
    \label{eq:eq7}
\end{eqnarray}
According to Eqs.~(\ref{eq:eq7}), the parameter $a$ has a twofold influence on the characteristics of the model: (i) it makes the proposed model non-local 
and (ii) is provides a kind of spatial scale of the system.
These two crucial properties are illustrated in Fig.~(\ref{fig:fig1}) which provides a graphical sketch of the proposed model.
As seen from Fig.~(\ref{fig:fig1}), the proposed -- continuous in space -- model is a superposition of an uncountable infinite set of discrete SSH-like models which contain a local hopping $v$ between the upper and the lower component of the spinor field $\mathbf{\Psi}(x)$ as well as a non-local hopping $w$ between the upper component of $\mathbf{\Psi}(x)$ and the lower component of $\mathbf{\Psi}(x-a)$. These SSH-like models become interconnected after imposing the assumption that $\mathbf{\Psi}$ is a smooth function of $x$.

To find the spectrum of $\mathbf{H}$ one assumes plane wave solutions of the form
\begin{eqnarray}
\Psi_A(x)&=&A_+ e^{\mathfrak{i} k x} + A_- e^{-\mathfrak{i}kx} \nonumber 
\\
\Psi_B(x)&=&B_+ e^{\mathfrak{i}kx} + B_- e^{-\mathfrak{i}kx}
\label{eq:eq8}
\end{eqnarray}
Inserting this plane wave ansatz into Eqs.~(\ref{eq:eq7}) leads to the homogeneous linear system of equations:
\begin{eqnarray}
(v + w e^{\mathfrak{i}ka}) A_+ - E B_+ = 0 \nonumber \\
(v + w e^{-\mathfrak{i}ka}) A_- - E B_- = 0 \nonumber \\
(v + w e^{-\mathfrak{i}ka}) B_+ - E A_+ = 0\\ \nonumber
(v + w e^{\mathfrak{i}ka}) B_- - E A_- = 0
\label{eq:eq9}
\end{eqnarray}
for $A_{\pm}$, $B_{\pm}$, which has a nonzero solution only if the determinant of the corresponding coefficients is zero. This in turn leads to the eigenvalue spectrum of $\mathbf{H}$:
\begin{equation}
    E_{\pm}(k)=\pm \sqrt{v^2+w^2+2 v w \cos{ak}}
    \label{eq:eq10}
\end{equation}
which constitutes a periodic band structure with two bands, symmetric around $0$ and with period $\displaystyle{{2 \pi \over a}}$. 
For $a=1$ the spectrum in Eq.~(\ref{eq:eq10}) coincides with the bulk spectrum of the corresponding discrete infinite-lattice SSH.  
This is expected according to the representation of $\mathbf{H}$ in Eq.~(\ref{eq:eq3}). The associated normalized eigenvectors are given as: 
\begin{eqnarray}
\mathbf{\Psi}_{k}^{(\pm)}(x) = \vert u_{k}^{(\pm)} \rangle \frac{e^{\mathfrak{i} k x}}{\sqrt{2 \pi}}~~~\nonumber \\
\mathrm{with}~~~|u_{k}^{(+)}\rangle = \frac{1}{\sqrt{2}} \begin{pmatrix} \displaystyle{e^{\mathfrak{i} \phi(k)/2}} \\ \displaystyle{e^{-\mathfrak{i} \phi(k)/2}} \end{pmatrix} \nonumber \\
|u_{k}^{(-)}\rangle = \frac{1}{\sqrt{2}} \begin{pmatrix}  \displaystyle{e^{\mathfrak{i} \phi(k)/2}} \\
-\displaystyle{e^{-\mathfrak{i} \phi(k)/2}} \end{pmatrix}
\label{eq:eq11}
\end{eqnarray}
where the phase $\phi(k)$ is: 
\begin{eqnarray}
\phi(k)=\arctan\left(\displaystyle{\frac{-w \sin ka}{v+w \cos ka}}\right) \nonumber \\ + \pi \;\Theta(-(v+w \cos ka)) \;\mathrm{sgn}(-w \sin ka)
\label{eq:eq12}
\end{eqnarray}
with $\Theta(x)$ being the Heaviside step function with the convention $\Theta(0)=1$ and $\mathrm{sgn}(x)=2 \Theta(x)-1$. In Eqs.~(\ref{eq:eq11}) the symbol $\pm$ is used to indicate the eigenvectors to $E_+(k)$ (symbol $+$) and $E_-(k)$ (symbol $-$) (as given in Eq.~(\ref{eq:eq10})) respectively. The normalization used in Eq.~(\ref{eq:eq11}) is:
\begin{eqnarray}
    \displaystyle{\int_{-\infty}^{\infty}}\; dk \;\;(\mathbf{\Psi}_{k}^{(\pm)}(x))^{\dagger} \mathbf{\Psi}_{k}^{(\pm)}(x^{\prime}) = \delta(x-x^{\prime})
\nonumber \\	\displaystyle{\int_{-\infty}^{\infty}}\; dx \;\;(\mathbf{\Psi}_{k}^{(\pm)}(x))^{\dagger}(x) \mathbf{\Psi}_{k^{\prime}}^{(\pm)}(x) = \delta(k - k^{\prime}) 
    \label{eq:eq13}
\end{eqnarray}

One can also calculate the Zak phase associated with each band:
\begin{equation}
\gamma_{\pm}=\displaystyle{\int_{-\frac{\pi}{a}}^{\frac{\pi}{a}}} \mathrm{i} \langle u_{k}^{(\pm)}\vert \partial_k u_{k}^{(\pm)} \rangle~dk = \begin{cases} \phantom{\pm}0,~~~\mathrm{for}~~ \vert v \vert  >  \vert w \vert \\ \pm \pi,~~~\mathrm{for}~~\vert v  \vert <  \vert w \vert \end{cases}
    \label{eq:eq14}
\end{equation}
which is exactly the behavior of the standard discrete SSH model. Thus, according to Eq.~(\ref{eq:eq14}), the non-local Hamiltonian $\mathbf{H}$ introduced in Eq.~(\ref{eq:eq1}) describes bulk topological insulator properties in continuous $x$-space while sharing key features with its discrete analogue. Based on $\mathbf{H}$ one can derive a family of models in a continuous $x$-space, which allows the development of a systematic path connecting local with non-local description. This is the main subject of the next section. 

\section{From local to non-local SSH in continuous space}

In this section, our strategy is to expand the translation operator $\mathbf{T}(a)$ in Eq.~(\ref{eq:eq4}) in powers of $a \frac{d}{dx}$ keeping terms up to order $O((a \frac{d}{dx})^n)$. Thus, for increasing $n$, we obtain a more accurate local approximation $\mathbf{H}^{(n)}$ of the Hamiltonian $\mathbf{H}$ in Eq.~(\ref{eq:eq1}). Clearly, in the limit $n \to \infty$ we recover the non-local Hamiltonian $\mathbf{H}$. In fact, as $n$ increases, higher and higher derivative terms appear in the off-diagonal elements of $\mathbf{H}^{(n)}$, making the corresponding approximate Hamiltonian less and less local. We will consider here the cases $n=0,~1$ and $2$ focusing on the bulk spectrum and its topological characteristics. Notice that for each obtained model Hamiltonian $\mathbf{H}^{(n)}$ the associated spectrum is calculated exactly, without assuming that $a$ is a small parameter.

 For $n=0$ we have $\mathbf{T}(a)=1$ and the Hamiltonian simplifies to $\mathbf{H}^{(0)}=(w+v) \bm{\sigma}_x$. This leads to a spectrum with two infinitely degenerate flat bands with eigenvalues $E^{(0)}_{\pm}=\pm (w+v)$. There is a continuous set of plane wave eigenvectors with $k \in (-\infty,\infty)$. Without extra assumptions, there is no way to obtain any non-trivial topological property in this case. Thus, we proceed to the case $n=1$. The translation operator becomes $\mathbf{T}(a)=1 + a \displaystyle{d \over dx}$ and the corresponding Hamiltonian $\mathbf{H}^{(1)}$ becomes:
\begin{equation}
\mathbf{H}^{(1)}=\begin{pmatrix}
    0 & w + v - w a\frac{d}{d x} \\ w + v + w a \frac{d}{d x} & 0
\end{pmatrix}
    \label{eq:eq15}
\end{equation}
with the energy spectrum:
\begin{equation}
E_{\pm}^{(1)}(k)=\pm\sqrt{(w+v)^2+(a k w)^2}
    \label{eq:eq16}
\end{equation}
and normalized eigenstates given by:
\begin{eqnarray}
\mathbf{\Psi}_k^{(\pm,1)}(x) = \vert u_k^{(\pm,1)} \rangle \frac{e^{\mathfrak{i} k x}}{\sqrt{2 \pi}} \nonumber \\
\mathrm{with}~~~|u_k^{(\pm,1)}\rangle = \frac{1}{\sqrt{2}} \begin{pmatrix} 1 \\ \displaystyle{\frac{v + w + 
\mathfrak{i} a k w}{E_{\pm}^{(1)}(k)}} \end{pmatrix}
\label{eq:eq17}
\end{eqnarray}
The superscript "$+$" ("$-$") refers to the positive (negative) energy branch, given in Eq.~(\ref{eq:eq16}), respectively, while the subsequent index $1$ is used to indicate that we are considering the Hamiltonian $\mathbf{H}^{(1)}$. Notice that $k \in (-\infty,\infty)$. Since the spectrum is not periodic with respect to $k$ the Zak phase cannot be strictly defined for this system. One can try to apply an one-point compactification of the $k$-space at $k=\pm \infty$, however this requires that the Hamiltonian $\mathbf{H}^{(1)}$ expressed in momentum basis fulfills: $\displaystyle{\lim_{k \to \infty}} \mathbf{H}^{(1)}(k)=\displaystyle{\lim_{k \to -\infty}} \mathbf{H}^{(1)}(k)$ which is not the case here (in fact it holds $\displaystyle{\lim_{k \to \infty}} \mathbf{H}^{(1)}(k)=\displaystyle{\lim_{k \to -\infty}} (\mathbf{H}^{(1)})^T(k)$).
Nevertheless, one can calculate the integral of the Berry connection along each band. This leads to:
\begin{equation}
\gamma^{(1)}_{\pm}=\displaystyle{\int_{-\infty}^{\infty}} \mathfrak{i} \langle u_k^{(\pm,1)}\vert \partial_k u_k^{(\pm,1)} \rangle~dk = - \frac{\pi}{2} \mathrm{sgn}(a w (v+w)) 
 \label{eq:eq18}
\end{equation}
Notice that the result obtained in Eq.~(\ref{eq:eq18}) is gauge dependent. Similar behavior for the integral of the Berry connection is also met in the Jackiw-Rebbi and related models \cite{Jackiw1976}. This indicates that the Hamiltonian $\mathbf{H}_c^{(1)}$ can support a topological interface mode by breaking the translational symmetry through spatially non-uniform $v(x)$ or/and $w(x)$ possessing a domain wall profile.

\begin{figure}[htp]
\begin{center}
\includegraphics[width=0.5\textwidth]{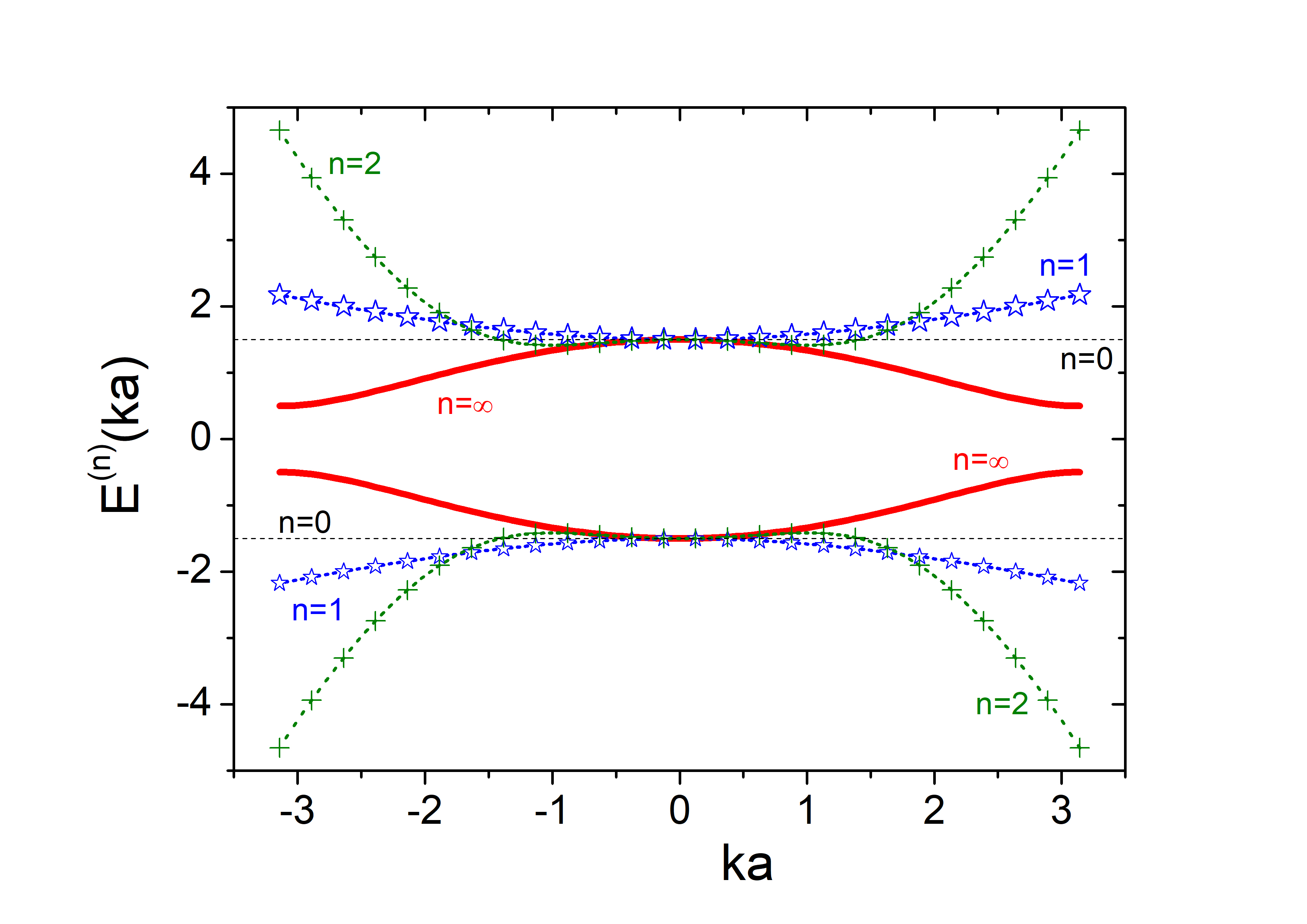}
\caption{The spectrum of the non-local continuous model for $w=1$ and $v=0.5$ (in a.u.) shown with red solid lines. We also show the spectra for its approximations obtained through the expansion of the translation operator up to $O(a^n)$ with $n=1$ (blue stars connected with a blue dashed line) and $n=2$ (olive crosses connected with an olive dashed line). Only the region $k a \in [-\pi,\pi]$ is displayed. The horizontal black dashed lines present the infinitely degenerate flat bands, occurring for $n=0$. }
\label{fig:fig2}
\end{center}
\end{figure}

Continuing along this line, we use the translation operator approximation $\mathbf{T}(a)=1 + a + \displaystyle{\frac{a^2}{2} \frac{d^2}{dx^2}}$ obtained for $n=2$, to determine the Hamilton operator $\mathbf{H}^{(2)}$:
\begin{eqnarray}
\mathbf{H}^{(2)}= 
\begin{pmatrix}
	0 & h_{-}\\ 
	h_{+} & 0
\end{pmatrix}
    \label{eq:eq19}
\end{eqnarray}
with $h_{\pm} = w + v \pm w a\frac{d}{d x} + w \frac{a^2}{2} \frac{d^2}{d x^2}$.
The corresponding energy spectrum is the following:
\begin{equation}
E_{\pm}^{(2)}=\pm \sqrt{\left( w + v -\frac{w}{2}(a k)^2\right)^2+(a w k)^2}
    \label{eq:eq20}
\end{equation}
Before proceeding to calculate the associated eigenvectors, it is worth discussing, at this point, the variation of the energy spectrum as the order $n$ of the local approximation to the translation operator in $\mathbf{H}^{(n)}$ increases. To this end, based on Eqs.~(\ref{eq:eq10},\ref{eq:eq16},\ref{eq:eq20}) one can graphically present the eigenvalue spectrum of $\mathbf{H}^{(n)}$ for $n=0$, $1$, $2$ compared to the spectrum of $\mathbf{H}$, illustrating the transition from a local to a non-local Hamiltonian description. The result is displayed in Fig.~\ref{fig:fig2}. The parameter values used in the figure are $w=1$ and $v=0.5$ in arbitrary units (a.u.). For finite-order approximations truncating the expansion at order $n$ in powers of $a$ ($n=0,1,2$), the spectrum is continuous with a gap size of $2\vert w+v \vert$ at $ka=0$. The spectra for $n=1$ and $n=2$ are plotted as blue stars (dashed blue line) and olive crosses (dashed olive line), respectively. The figure clearly indicates, for increasing $n$, the progressive proximity toward the spectrum  of the non-local model (solid red lines) near $ka=0$. Crucially, in these local models the gap could close only when $v=-w$ at $k a=0$, indicating the need for a negative hopping amplitude.  In contrast, the exact non-local model ($\mathbf{H}$) has a gap closing both when $w v >0$ (at $ka=\pm \pi$) and when $w v <0$ (at $ka=0$).  The horizontal black dashed lines in Fig.~(\ref{fig:fig2}) present the discrete spectrum of $\mathbf{H}^{(0)}$, appearing for $\mathbf{T}(0)=1$, with infinitely degenerate flat bands, as previously discussed.

Let us now proceed with the determination of the eigenstates of $\mathbf{H}^{(2)}$. We obtain:
\begin{eqnarray}
\mathbf{\Psi}_k^{(\pm,2)}(x) = \vert u_k^{(\pm,2)} \rangle \frac{e^{\mathfrak{i} k x}}{\sqrt{2 \pi}} \nonumber \\
|u_k^{(\pm,2)}\rangle = \frac{1}{\sqrt{2}} \begin{pmatrix} 1 \\ \displaystyle{\frac{v + w  - \frac{w}{2}(a k)^2 + \mathfrak{i} a k w}{E_{\pm}^{(2)}(k)}} \end{pmatrix}
\label{eq:eq21}
\end{eqnarray}

The associated integral of the Berry connection for this model leads to
\begin{equation}
\gamma^{(2)}_{\pm}=- \pi \Theta(w) 
\label{eq:eq22}
\end{equation}
allowing for the emergence of topologically protected localized boundary modes when $w > 0$. Notice that in this case, the one-point compactification at $k=\pm \infty$ is possible since $\displaystyle{\lim_{k \to \infty}} \mathbf{H}^{(2)}(k)=\displaystyle{\lim_{k \to -\infty}} \mathbf{H}^{(2)}(k)$. Thus, $\gamma^{(2)}_{\pm}$ can be considered as a generalized Zak phase. In this system, a topological transition occurs when the gap of the bulk spectrum closes at $k a=0$ under the condition that $v=-w$, indicating that also in this model a negative hopping amplitude is needed
for its occurrence. A final remark is here in order: all Hamiltonians $\mathbf{H}^{(n)}$ which provide a local approximation to $\mathbf{H}$ respect the chiral symmetry. Thus, any topological state occurring in these systems is expected to be robust against disorder which does not violate the chiral symmetry. A detailed discussion of the topological characteristics of the $n$-th order local approximations $\mathbf{H}^{(n)}$ to $\mathbf{H}$ goes beyond the scope of the current work. Instead, in the next section we will focus on the formulation of a finite-size version of the non-local model searching for the emergence of topological edge modes.

\section{Zero energy topological flat mini band in the finite non-local SSH model in continuous space}

\begin{figure}[htp]
\begin{center}
\includegraphics[width=0.5\textwidth]{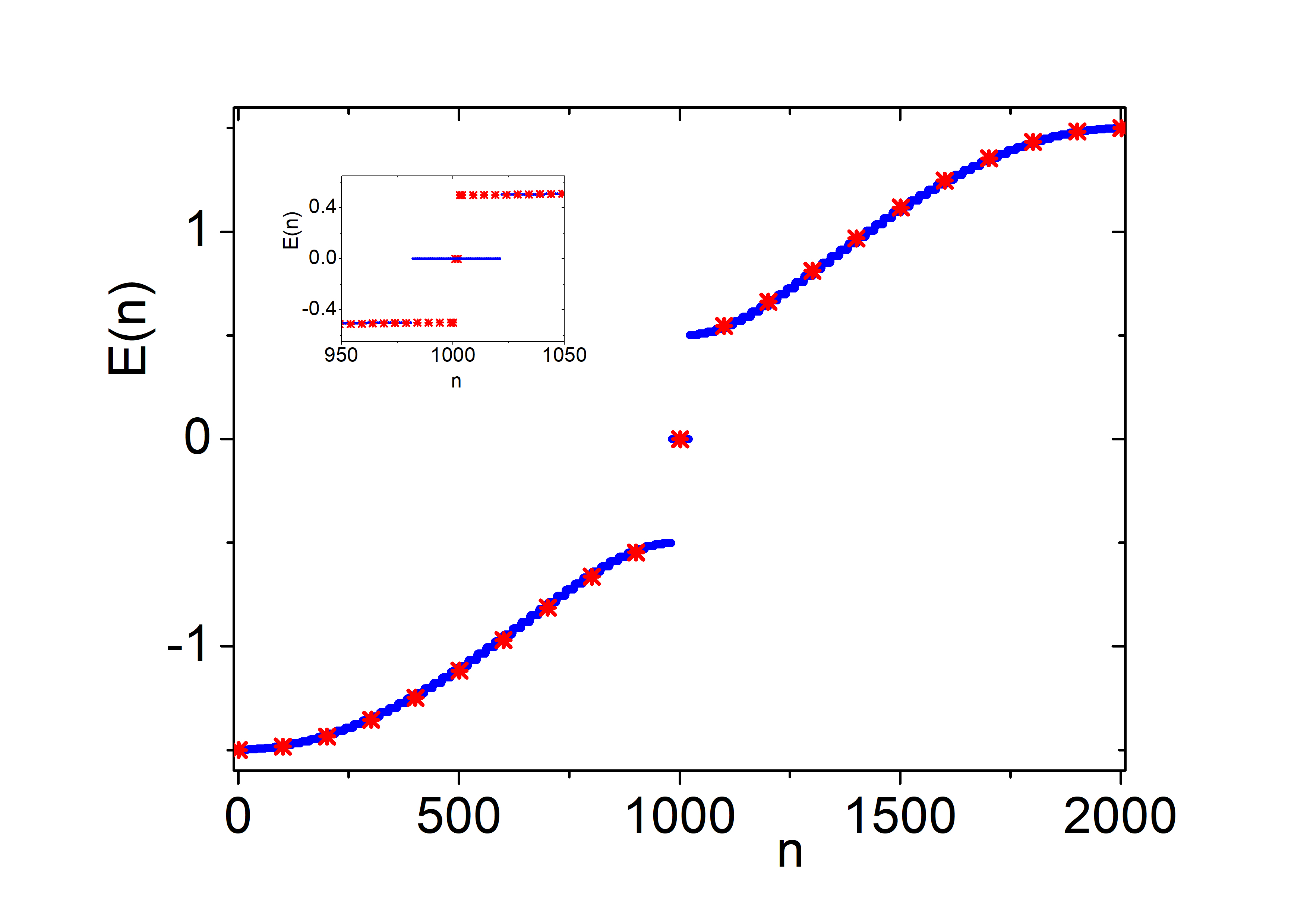}
\caption{The spectrum of the finite non-local, continuous in space, model $\mathbf{H}_{L}$ for $w_0=1$ and $v_0=0.5$ (in a.u.) shown with blue solid lines. In the same plot we display with red crosses the spectrum of the discrete SSH lattice for the same values of $w_0$, $v_0$, open boundary conditions and number of sites $N=2002$. For illustrative reasons we show every $100$-th eigenenergy of the discrete system. In the inset we magnify the region around $\mathcal{E}_{L}=0$. There is an accumulation of zero energy states in the non-local model in contrast to the discrete SSH lattice where only two states with energy close to zero appear.}
\label{fig:fig3}
\end{center}
\end{figure}

The non-local character of $\mathbf{H}$, as expressed in Eq.~(\ref{eq:eq4}), makes the formulation of a finite system based on this Hamiltonian a non-trivial task. The eigensystem relations in Eq.~(\ref{eq:eq7}) connect the components of the spinor field at point $x$ with those at points $x\pm a$. Thus, when $x$ is at the boundary region of the finite system, it requires some care for the treatment of these equations to avoid mathematical inconsistencies. In fact, the finite non-local system cannot be considered without reference to its environment. In the simplest scenario, the finite one dimensional non-local SSH of size $L$ should be treated as a system defined in a finite region $[-{L \over 2}, {L \over 2}]$ of the entire $x$-space. Since translational symmetry is broken in this case, one expects that the parameters $w$, $v$ of the model are functions of the position $x$, as they must vanish outside the system. Therefore, we define the spatially varying parameters: 
\begin{eqnarray}
    v(x)=\begin{cases} v_L~\mathrm{for}~x \in [-{L \over 2}, {L \over 2}] \\
    0~~~\mathrm{for}~x \notin [-{L \over 2}, {L \over 2}] \end{cases} \nonumber \\
	\mathrm{and}~~w(x)=\begin{cases} w_L~\mathrm{for}~x \in [-{L \over 2}, {L \over 2}] \\
    0~~~\;\mathrm{for}~x \notin [-{L \over 2}, {L \over 2}] \end{cases}
    \label{eq:eq23}
\end{eqnarray}
with $v_L$, $w_L$ constant.
Promoting the parameter $w$ to a function $w(x)$ introduces a subtle issue: the ordering of factors in $w \mathbf{T}(\pm a)$ in Eq.~(\ref{eq:eq4}). It follows that, for the finite Hamiltonian $\mathbf{H}_{L}$ to be Hermitian, the translation operator must also act on the function $w(x)$ to its right, yielding $\mathbf{T}(\pm a)\left(w(x) ..\right)$.  This becomes evident upon considering the $O(a {d \over dx})$ expansion of $\mathbf{T}(\pm a)$ and imposing the Hermiticity of $\mathbf{H}^{(1)}$ in Eq.~(\ref{eq:eq15}), under the assumption of a spatially varying $w(x)$. Taking this into account, the eigenvalue problem for the Hamiltonian $\mathbf{H}_{L}$ leads to the equations:
\begin{eqnarray}
    w(x-a) \psi_B(x-a) + v(x) \psi_B(x) &=& \mathcal{E}_{L} \psi_A(x) \nonumber \\
    w(x+a) \psi_A(x+a) + v(x) \psi_A(x)&=& \mathcal{E}_{L} \psi_B(x)
    \label{eq:eq24}
\end{eqnarray}
where $\psi_a(x)$ with $a=A,B$ are the components of the eigenspinor of the finite Hamiltonian $\mathbf{H}_{L}$ and $\mathcal{E}_{L}$ is the corresponding energy eigenvalue. Eqs.~(\ref{eq:eq24}) can be solved numerically, leading to the spectrum given in Fig.~(\ref{fig:fig3}). 
The solid blue line is the spectrum of the finite system given by $\mathbf{H}_{L}$, while the red stars present the bulk spectrum of the corresponding finite SSH lattice with the same parameters $v_0$, $w_0$. The wiggle/plateau structure of the blue line is an artifact of discretization used in numerical calculations. For the latter, we have used $L=10$  $a=0.2$, $v_0=0.5$ and $w_0=1$ (all in a.u.). The space discretization has been performed using $\Delta x=0.01$ (a.u.). The number of sites in the discrete SSH lattice is taken as $N=2(\lfloor{ L \over \Delta x}\rfloor + 1)$. Thus, we conclude that the Hamiltonian of $\mathbf{H}_{L}$ has a bulk spectrum that is very close to that of a finite discrete SSH model. 
There is very minor deviations between the two spectra being invisible at the scale of the figure.
However, in the spectrum of $\mathbf{H}_{L}$ we observe an accumulation of states near zero energy, in contrast to the discrete SSH lattice, where only two such states appear. Furthermore, the numerical results for the spectrum of $\mathbf{H}_L$ indicate that the flat miniband shown in the inset of Fig.~(\ref{fig:fig3}) lies exactly at zero energy. This differs from the discrete SSH lattice, in which the states close to the midgap have energies that are approximately, but not exactly, zero and come in pairs with opposite signs. It should be also clarified that the comparison between the two models is made imposing hard-wall boundary conditions to the non-local continuous model and open boundary conditions to the discrete model. Assuming Dirichlet boundary conditions for the SSH lattice introduces even more discrepancies from the continuous model particularly in the profile of the associated eigenvectors.

Notice that, although the continuous non-local model can be interpreted as a superposition of an infinite and uncountable set of discrete SSH models (as previously discussed in connection with Fig.~(\ref{fig:fig1})), any spatial information regarding the position of these SSH lattices is meaningless unless they are explicitly embedded in the continuous $x$-space. Consequently, this interpretation of the continuous non-local model in terms of SSH lattices is limited and necessarily requires the involvement of continuous space. As a result, deviations such as the one shown in the inset of Fig.~(\ref{fig:fig3}) are expected to originate from this fact. Therefore, it is worthwhile to examine the states belonging to the zero-energy sector of $\mathbf{H}_{L}$.

For $\mathcal{E}_{L}=0$ the Eqs.~(\ref{eq:eq24}) decouple
and therefore can be solved straightforwardly. The general solution can be written as:
\begin{equation}
    \psi_s(x)=U_s(x) e^{q_s x}~~~~,~s=A,~B~~; U_s(x+a)=U_s(x)
    \label{eq:eq25}
\end{equation}
where $U_s(x)$ is a periodic function with period $a$ and $q_s$ has to be determined. Inserting the ansatz in Eq.~(\ref{eq:eq25}) into the Eqs.~(\ref{eq:eq24}) and using $\mathcal{E}_{L}=0$ we find:
\begin{eqnarray}
q_s=\eta_s \frac{1}{a} \ln \Big\vert \displaystyle{\frac{w_0}{v_0}} \Big\vert - \mathfrak{i} \eta_s \frac{\pi}{a} \nonumber \\
s=A,~B~~~\mathrm{and}~~~\eta_A=-1,~\eta_B=1
    \label{eq:eq26}
\end{eqnarray}
Thus, there is an exponential envelope for the $\psi_s(x)$ components of the zero energy eigenstates of $\mathbf{H}_{L}$ together with an oscillating part with period $\displaystyle{\frac{a}{2}}$, i.e. half the period of $U_s(x)$. The envelope of the $A$-component $\psi_A(x)$ decays exponentially as $x$ varies from $\displaystyle{-\frac{L}{2}}$ to $\displaystyle{\frac{L}{2}}$ while the $B$-component $\psi_B(x)$ has the opposite behavior (exponential growth as $x$ varies from $\displaystyle{-\frac{L}{2}}$ to $\displaystyle{\frac{L}{2}}$). The general form of the $a$-periodic part $U_s(x)$ in $\psi_s(x)$ can be chosen as:
\begin{equation}
   U_s(x)=\cos \left( \frac{2 \pi n_s}{a} x + \phi_{n_s} \right)~~~;~~~n_s \in \mathbb{N}  
    \label{eq:eq27}
\end{equation}
where the phase $\phi_{n_s}$ is determined by the boundary conditions. It is also worth to notice here that the scale of non-locality $a$ influences both the value of the localization length characterizing the exponential decay as well as the period of the oscillating part of these zero modes.

\begin{figure}[htp]
\begin{center}
\includegraphics[width=0.5\textwidth]{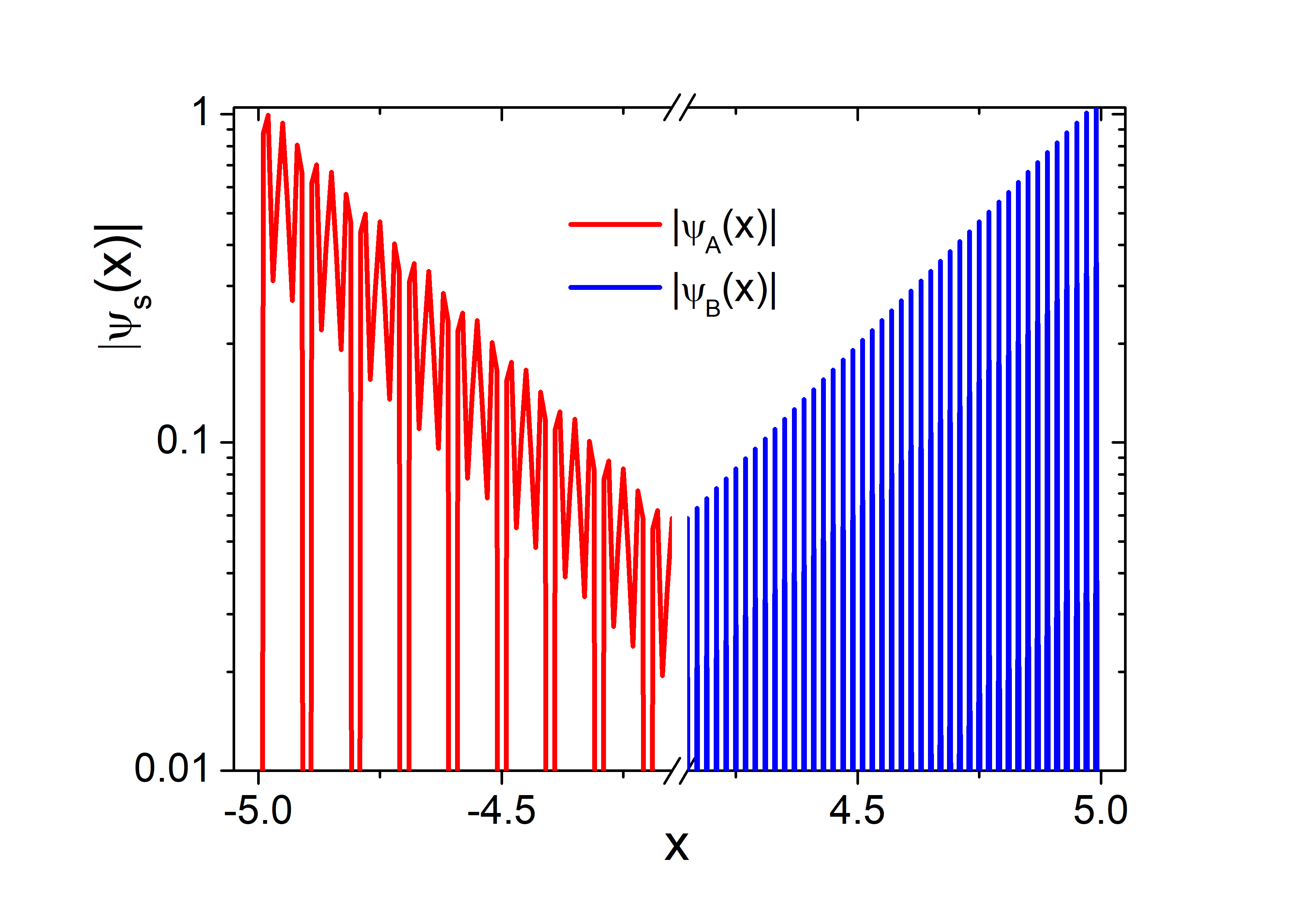}
\caption{The magnitude of the zero energy eigenstate-components $\psi_s(x)$ ($s=A$, $B$) with $(n_A,m_A)=(3,1),~(n_B,m_B)=(5,2)$ on a logarithmic scale. The parameter values used are $v_0=0.5$, $w_0=1$, $L=10$ and $a=0.2$ all in arbitrary units. The exponential envelope with characteristic exponent $\displaystyle{\frac{1}{a} \ln  \Big\vert \frac{w_0}{v_0} \Big\vert} \approx 1.5$ is clearly recognizable. Notice that $\psi_s(x) $ under the action of the operator $\bm{\sigma}_x \mathbf{\Pi}(x)$ (generalized parity) transforms to the function $\tilde{\psi}_s(x)$ with $\tilde{\psi}_A(x)=\psi_B(-x)$ and $\tilde{\psi}_B(x)=\psi_A(-x)$.}
\label{fig:fig4}
\end{center}
\end{figure}

For the assumed hard-wall boundary conditions we obtain:
\begin{equation}
    U_s \left( \displaystyle{-\frac{L}{2}} \right) = U_s \left( \displaystyle{\frac{L}{2}} \right) = 0~~~;~~~s=A,~B
    \label{eq:eq28}
\end{equation}
Eq.~(\ref{eq:eq28}) determines the phase $\phi_{n_s}$, up to an odd multiple of $\displaystyle{\frac{\pi}{2}}$, as:
\begin{eqnarray}
    \phi_{n_s}(m_s)=(2 m_s + 1) \frac{\pi}{2} + n_s \frac{\pi  L}{a} \nonumber \\
s=A,~B~~~\text{and}~~~n_s \in \mathbb{N},~m_s \in \mathbb{Z}
    \label{eq:eq29}
\end{eqnarray}
As a consequence, Eq.~(\ref{eq:eq29}) implies the existence of an infinite, countable set of eigenspinors to $\mathbf{H}_{L}$ classified by two integer numbers $(n_s,m_s)$ for each spinor component $\psi_A$, $\psi_B$. Both components are modulated by an exponential envelope leading to the localization to the left (right) in the interval $[-{L \over 2}, {L \over 2}]$ of the component $\psi_A(x)$ ($\psi_B(x)$) respectively. In Fig.~(\ref{fig:fig4}) we show the magnitudes $\vert \psi_A(x)\vert$, $\vert \psi_B(x) \vert$ for the zero energy eigenstate with $(n_A,m_A)=(3,1),~(n_B,m_B)=(5,2)$ on a logarithmic scale. A break on the $x$-axis is introduced to allow for better illustration of the functional behavior close to the boundaries.
Thus, we observe that the introduced non-local model defined in continuous $x$-space possesses an entire flat mini band at zero energy in its finite-size version. The corresponding eigenstates are spinors with components localized at the left (upper component) and right (lower component) edges of the system, respectively. Due to chiral symmetry and gap protection, it is expected that these states are robust against disorder which respects the chiral symmetry. Furthermore, the corresponding localization length is controlled by the parameter $a$ which quantifies the spatial scale and/or the non-locality of the system. 

\section{Concluding remarks}

We introduce a continuous, infinitely extended, non-local model that captures nearly all the phenomenological features of the discrete SSH model. It shares the same bulk energy spectrum and bulk eigenstates. The associated Zak phase also behaves identically to that of the SSH model, indicating a similar topological structure. The model includes a parameter $a$, with dimension length, which represents a spatial scale and is responsible for the models' non-locality. Notably, the energy spectrum and the corresponding eigenstates become exactly identical to those of the SSH model through appropriate tuning of this parameter.

In contrast to other continuous analogues of the SSH model based on Schr\"{o}dinger-type Hamiltonians, the proposed non-local model does not contain external potential terms and therefore preserves exact chiral symmetry. Moreover, unlike continuous SSH analogues derived from Dirac-type Hamiltonians, our model exhibits periodic energy bands.

The model also supports the emergence of a countable infinite set of exponentially localized zero-energy edge states when considered on a finite spatial domain. This set forms a mini flat band at zero energy in contrast to the discrete finite SSH model, where two nearly zero-energy edge states of opposite sign and location appear. Furthermore, the localization length of the edge states belonging to this flat band can be modified by varying the non-locality parameter $a$. In this respect, the model exhibits important differences from the discrete SSH model, which suggests also to revisit the formulation of bulk-edge correspondence.

Our method for deriving continuous analogues of bipartite lattices is quite general and can be extended to other bipartite systems beyond SSH, as well as to more complex multipartite lattices. Furthermore, it would be interesting to consider the possibility to formulate the model in two spatial dimensions and explore the eventual emergence of localized states there.

Finally, a note is in order concerning a corresponding experimental 
realization of our continuous non-local model.
The standard SSH model has been implemented and analyzed experimentally in a variety of
different physical situations and setups, such as the topological band structure of ultracold trapped atoms \cite{Cooper2019},
water waves in channels in the linear and nonlinear regime \cite{Anglart2025}, photonic superlattices \cite{Wang2021,Klauck2021}, and
mechanical spring-connected systems \cite{Thatcher2022}. Non-locality could possibly be added to some of these setups analogous
to the case of a helix, where long-range hopping across windings is enabled by the curved geometry \cite{Stockhofe2015}.
Still, whether our spatially non-local continuous space analogues of bipartite (or multipartite) lattices can
be realized experimentally remains an open and challenging question for future research.

\section{Acknowledgments}
The authors are very grateful to T. Posske for a very careful reading of the manuscript and several valuable suggestions.
This work has been supported by the Cluster of Excellence “Advanced Imaging of Matter” of the Deutsche
Forschungsgemeinschaft (DFG) - EXC 2056 - project ID 390715994.

\end{document}